\documentclass[aps,pre,onecolumn,showpacs]{revtex4}
\usepackage{graphicx}
\usepackage{wasysym}

\newcommand{\be}{\begin{equation}}
\newcommand{\ee}{\end{equation}}
\newcommand{\bea}{\begin{eqnarray}}
\newcommand{\eea}{\end{eqnarray}} 
\newcommand{\lanran}{\langle x(t)\rangle _T} 

\begin{document}

\title{Average trajectory of returning walks}

\author{Francesca Colaiori}
\email{fran@pil.phys.uniroma1.it}
\author{Andrea Baldassarri}
\email{andrea.baldassarri@roma1.infn.it}
\author{Claudio Castellano}
\email{castella@pil.phys.uniroma1.it}
\affiliation{Dipartimento di Fisica, Universit\`a di 
Roma ``La Sapienza'', and Istituto Nazionale per la Fisica della Materia,
Unit\`a di Roma 1, P.le A. Moro 2, I-00185 Roma, Italy}

\date{\today}

\begin{abstract}

We compute the average shape of trajectories of some one--dimensional
stochastic processes $x(t)$ in the $(t,x)$ plane during an excursion,
i.e. between two successive returns to a reference value, finding
that it obeys a scaling form.  For uncorrelated random walks the
average shape is semicircular, independently from the single
increments distribution, as long as it is symmetric.  
Such universality extends to biased random
walks and Levy flights, with the exception of a particular class of
biased Levy flights.  Adding a linear damping term destroys scaling
and leads asymptotically to flat excursions.  The introduction of
short and long ranged noise correlations induces non trivial
asymmetric shapes, which are studied numerically.

\end{abstract}

\pacs{05.40.-a,75.60.Ej,05.45.Tp}

\maketitle

\section{Introduction}
\label{Introduction}

Many disordered systems respond to external solicitations by 
producing noise with power law features, that can be modeled 
in terms of avalanches. A notable example of such phenomena is 
the Barkhausen effect, first observed about a century ago by recording 
the noise produced by the reversal of large domains in a ferromagnet. 
The Barkhausen noise has been incessantly investigated, both because of 
its practical application as a nondestructive method to test magnetic 
materials, and because of its conceptual relevance for the understanding 
of the magnetization dynamics on a microscopic scale \cite{bar}. 
Experiments show that both the size and the duration of avalanches of spin 
reversal are power law distributed over several decades. The exponents 
characterizing these power laws are often used to identify universality 
classes~\cite{zap,kun}.
Recently, the average pulse shape has been proposed as a sharper tool 
for discriminating among universality classes and to test models against 
experiments~\cite{nat}. 
This analysis has revealed some weaknesses of present models, 
since they all fail to reproduce the avalanche shapes observed experimentally. 
Namely, all models proposed so far produce symmetric shapes, 
while leftward skewed forms are observed in experiments, indicating that our
understanding of the Barkhausen effect is, at the present stage, incomplete. 
This open issue has been the inspiration of this work.

We consider the problem of finding the average shape of a generic
stochastic signal during an excursion, that is between two successive
returns to a reference value.
We hope that a deeper understanding of how the statistical properties 
of the signal are reflected on the shape of the average excursion
can in general give insight into the understanding of the process
generating the signal. 
In the case of Barkhausen noise, this may help identifying which crucial 
ingredient is missing in the theory, and lead to the introduction of
more accurate models. 

Beyond its interest for what concerns the understanding of Barkhausen
noise, the nontrivial phenomenology of the avalanche shape leads to
more general and interesting questions: What are the physical
ingredients that determine the shape of the average excursion in a
generic stochastic process?  Are there universality classes?  Does
this shape encode pieces of information about the underlying physical
system, that are not accessible by considering other observables?
These issues have not been addressed so far.  In this paper we begin a
systematic investigation of the shape of the average excursion, by
considering some simple stochastic processes, both uncorrelated and
correlated, and with generically distributed increments.  In this way
we provide a first theoretical framework that may be of help in the
analysis of real time series in many contexts.

In Section~\ref{Definition} we introduce the concept of excursion, the
types of processes that we will consider in the following
and the general scaling form of the average excursion.
Section~\ref{Uncorrelated} presents the results for 
processes with uncorrelated increments
(Brownian motion, Random walk, Levy flights)
and Section~\ref{Damped} discusses the effect of a damping term in
a Brownian motion.
Sections~\ref{Long-range} and~\ref{Short-range} consider, respectively,
the effect of long- and short-ranged noise correlations.
The final Section~\ref{Conclusions} presents some conclusions
and an outlook.
A short account of some of the results presented here appeared in
Ref.~\cite{Baldassarri03}.

\section{Definition of the average excursion}
\label{Definition}
\begin{figure}
\includegraphics[angle=0,width=9cm,clip]{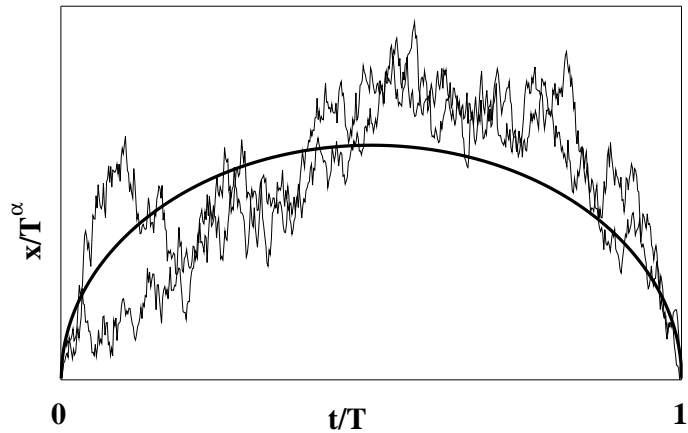}
\caption{Schematic representation of the average shape of a fluctuation.
Thin solid lines are two realizations of the stochastic process $x(t)$,
both returning for the first time at zero at time $T$. The thick solid
line is the average shape computed over many realizations.}
\label{Fig0}
\end{figure}

Let us first define the average excursion of a stochastic process.  We
consider a real valued $1d$ process $x(t)$ defined by a Langevin equation
with suitable initial conditions. An excursion of the process is the
trajectory in the $(t,x)$ plane, followed until the first return to the 
initial value $x(0)$ [see Fig.~(\ref{Fig0})].
We are interested in the statistics of positive excursions of a given
duration $T$, i. e. those such that $x(t)>x(0)=0$ for $0 < t< T$.
In particular we will denote the average excursion as $\lanran$.

~(\ref{scaling})
When analyzing real experimental data, one may need to extend our
definition to a generic reference value different from zero.  For
example, this is the case for positive signals, for which the
identification of excursion (avalanches) is made unclear by the
presence of background noise: one has to choose a recipe to decide
when an avalanche starts or ends. In practice, one sets a threshold
which is small enough not to change the shape of the avalanche, but
high enough with respect to background noise.

A generic reference value $a$ is taken into account by a translation
of the origin of $t$ and $x$: the excursion of $x(t)$ with respect to
$a$ is the excursion of a process $x'(t')=x(t'+t_a)-a$ with respect to
$x'(0)=0$, where $x(t_a)=a$.

Notice that in general the probability distribution of
first return times $P(T)$ depends on the threshold value $a$.
For this reason we expect that variations of $a$ will affect
the average excursion $\lanran$.
In fact, for most of the simple processes discussed below the choice of
$a$ has no effect on the shape of the average fluctuation.

We will analyze two kinds of processes.
The first are processes of the type
\be
\partial_t x(t)=\xi(t)
\ee
with random increments $\xi(t)$ extracted from a distribution $Q(\xi)$.
In Section~\ref{Uncorrelated} the noise $\xi$ will be taken to be
uncorrelated while in Sections~\ref{Long-range} and~\ref{Short-range}
we will deal with the effect of correlations.
The case considered in Section~\ref{Damped} is the simplest instance of a
wide class of processes, Brownian motions in a potential
\be
\partial_t x(t)=- dV(x)/dx + \xi(t)
\label{potential}
\ee

For analytical calculations it is useful to express the average
excursion more explicitly.  Let us first introduce the excursion
distribution $\Omega(x,t \mid x_0,0;x_0,T)$, which is the probability that a
trajectory, started in $x_0$ for $t=0$ and returning to $x_0$ for the
first time at time $T$, is in $x$ at time $t$.  For each time $t$,
$\Omega$ is the distribution of the quantity whose average is
$\lanran$, i. e. 
\be
\lanran = {\int_0^\infty dx \, x \,
\Omega(x,t \mid x_0,0;x_0,T) \over \int_0^\infty dx \,
\Omega(x,t \mid x_0,0;x_0,T)}
\ee
Note that
$\Omega(x,t \mid x_0,0;x_0,T)$ is related to the distribution of first
return times of
the process $P(T)\equiv \int_0^\infty dx \, \Omega(x,t \mid x_0,0;x_0,T)$.

In the case of Markovian processes, as Eq.~(\ref{potential}), 
$\Omega$ can be written in terms of the function
$c(x, t \mid x_0,t_0)$, which is the probability that the process started at
$x_0$ at time $t_0$ is in $x$ at time $t$, with the condition that
$x>0$ for all $t_0< t'<t$.
The probability $\Omega(x,t \mid x_0,0;x_0,T)$ of the whole trajectory
is the product
of the probability $c(x,t \mid x_0,0)$ of going from $x_0$ to $x(t)$ times
the probability $c(x_0,T \mid x,t)$ to go from $x(t)$ back to $x_0$ at time $T$.
Notice that instead of starting exactly from $0$ we have to consider the
path starting and arriving in $x_0 \to 0$ since $c$ vanishes identically
for $x_0=0$. Hence, for any Markovian process, we can write
\be
\lanran = \lim_{x_0 \to 0^+}
\frac{\int_0^\infty dx \, c(x,t \mid x_0,0) \, x \, c(x_0,T \mid x,t)}
{\int_0^\infty dx \, c(x,t \mid x_0,0) \, c(x_0,T\mid x,t)} \,.
\label{xbm}
\ee
Eq.~(\ref{xbm}), together with translational invariance and the
scaling assumption
\be
c(x,t \mid x_0,0) = t^\beta h((x-x_0)/t^\alpha)
\label{ansatz}
\ee
for the conditional probability $c(x,t \mid x_0,0)$ implies
\be
\lanran = T^\alpha f(t/T).
\label{scaling}
\ee
In some of the cases that we will consider Eq.~(\ref{xbm})
cannot be applied, since noise correlations break the Markovian property.
Nevertheless we will always find Eq.~(\ref{scaling}) to be true provided
that the distribution $P(T)$ of first return times decays algebraically.
In all the cases considered the exponent $\alpha$ coincides with
the wandering exponent of the unconstrained process, defined by 
$\langle [x(t)-x(0)]^2\rangle \simeq t^{\alpha}$.

\section{Uncorrelated processes}
\label{Uncorrelated}

\subsection{Brownian motion}
The simplest process is the uncorrelated Brownian motion 
\begin{equation}
\partial_t x(t)=\xi(t)
\label{bm}
\end{equation}
where $\xi(t)$ is a Gaussian white noise with $\langle\xi(t)\rangle=0$ and 
$\langle \xi(t)\xi(t')\rangle=\delta(t-t')$. 

The probability $c(x,t \mid x_0,0)$ can be computed via the image
method~\cite{Rednerbook} as a linear combination of two solutions
of the Fokker-Planck equation associated to the free process
\be
c(x,t \mid x_0,0) = \frac{2}{\sqrt{2\pi t}}
\left[ e^{-(x-x_0)^2/(2t)} - e^{-(x+x_0)^2/(2t)}
\right] \,,
\ee
yielding, in the limit of small $x_0$
\be
c(x,t \mid x_0,0) = \frac{2}{\sqrt{2\pi}t^{3/2}} x x_0 e^{-x^2/(2t)}.
\label{c}
\ee
The time-reversal and time-translational invariance of the process
imply $c(x_0,T \mid x,t)=c(x, T-t \mid x_0,0)$, hence the distribution
$\Omega$ is
\be
\Omega(x,t \mid x_0,0;x_0,T)= c(x,t \mid x_0,0) c(x, T-t \mid x_0,0) \propto
[(T-t) t]^{-3/2} (x x_0)^2 e^{-x^2 \left\{ 1/(2t)+1/[2(T-t)] \right\}}
\ee

Expression~(\ref{c}), inserted into Eq.~(\ref{xbm}), gives for the average
excursion
\be
\lanran =
T^{1/2} \sqrt{\frac{8}{\pi}} \sqrt{{t \over T} 
\left (1-{t \over T}\right )} \,.
\label{sc}
\ee
The average excursion of Brownian motion is thus of the scaling
form~(\ref{scaling}), with the exponent $\alpha=1/2$ coinciding
with the wandering exponent of the free process and a scaling
function proportional to a semicircle
\be
f_U(s) = \sqrt{8 \over \pi} \sqrt{s(1-s)},
\label{semicircle}
\ee
where $s=t/T$.
This result had already been noticed by Fisher~\cite{Fisher}.
The variance of the excursion is also readily computed
\be
\langle (x - \lanran)^2 \rangle_T = T \left( 3- {8 \over \pi} \right) s  (1-s).
\ee

The previous results are easily generalized to the case of a
Brownian motion with bias, that is Eq.~(\ref{bm}) with
$v \equiv \langle \xi \rangle > 0$.
The process is now invariant under time-reversal only provided the
velocity is also reversed.
Therefore in this case $c(x_0,T \mid x,t ;v)=c(x, T-t \mid x_0,0;-v)$.
Such quantities can again be computed via the image method~\cite{note1}
\begin{equation}
c(x,t \mid x_0,0;v)
= \frac{1}{\sqrt{2\pi t}} e^{-(x-x_0-vt)^2/(2t)} 
\left(1-e^{2 x x_0/t} \right)\,.
\end{equation}
Inserting this expression into the formula for the excursion distribution
it turns out that $\Omega$ is the same of the unbiased case except
for an additional factor $e^{-v^2 T/2}$.
Such a constant appears both in the numerator and the denominator
of Eq.~(\ref{xbm}) implying that
the average excursion is exactly the same of the unbiased Brownian motion
(\ref{sc}).

Notice that the addition of a bias introduces a characteristic time of order
$1/v^2$, which reflects in a cutoff in the distribution of first return
times~\cite{Feller}
\be
P(T) \propto T^{-3/2} e^{-v^2 T/2}.
\ee
However, a bias does not alter the shape of the excursion:
the number of trajectories that survive up to a time $T \gg 2/v^2$
is exponentially small, but the average shape of these unlikely
events is exactly the same as for the unbiased case.

This is the first example of the insensitivity of $\lanran$ with respect
to changes in the distribution of single steps, a feature that will turn out
to be quite generic.

\subsection{Random walk}
The Brownian motion is a continuous process in space and time.
On the basis of the central limit theorem it is reasonable to expect the
form of the average excursion to be the same for all processes
with finite variance of the single increments.
To support this conjecture we now compute $\lanran$ for
a process with finite variance, discrete in space and time, 
a random walk with bimodal distribution of the noise, i. e.
$Q(\xi) = (\delta_{\xi,1} + \delta_{\xi,-1})/2$.
The number of paths starting in $0$ at time $0$ and ending in
$x$ at time $t$ without  ever touching the $x=0$ axis is given
by~\cite{Rednerbook}
\begin{equation}
F(x,t)=\frac{x}{t}
\frac{t!}
{\left(\frac{t+x}{2}\right) !\left(\frac{t-x}{2}\right) ! }
\label{grwf}
\end{equation}
Hence the probability to find the walker in $x$ at time $t$ with the condition
that it has never touched the axis is obtained from Eq.~(\ref{grwf})
dividing $F(x,t)$
by $M(t)=\sum_x F(x,t)$, the total number of possible trajectories of
$t$ steps in the positive $x$ half-plane
\begin{equation}
c(x,t)=\frac{F(x,t)}{M(t)} \,.
\end{equation}
Using time reversal symmetry,
the average excursion is then given by
\begin{equation}
\lanran = \frac{\sum_x \, x \, c(x,t) \, c(x,T-t)}
{\sum_x \, c(x,t) \, c(x,T-t)} = K \sum_{x=0}^t x \, c(x,t) \, c(x,T-t) \,, 
\label{discretexbm}
\end{equation}
with $K=M(t)M(T-t)/F(1,T-1)$, where we used the fact that
$\sum_x F(x,t) F(x,T-t)$ is independent from $t$ and equal to $F(1,T-1)$.
Introducing the variables $\gamma = x/\sqrt{t}$, $s = t/T$, and
$\phi=\sqrt{s/(1-s)}$, and using the expansion $n!/(n/2)!\simeq \sqrt{2/\pi}
\, 2^n/n$, we get $M(t) c(\gamma \sqrt{t},t) = F(\gamma \sqrt{t},t) \simeq
\frac{2^t}{\sqrt{\pi}t} \gamma e^{-\gamma^2/2}$, 
$F(1,T-1)\simeq \frac{2^T}{\sqrt{2\pi} T^{3/2}}$ and
$K \simeq \sqrt{\frac{2}{\pi}}\frac{T^{-1/2}}{s(1-s)}$
so that the average excursion is
\begin{equation}
\lanran =
K t \phi \int_0^{\sqrt{t}} d\gamma \gamma^3 e^{-\frac{\gamma}{2}(1+\phi^2)}
\simeq \frac{2 K \phi t}{(1+\phi^2)^2}
\end{equation}
where in the last step the integral has been extended to infinity.
Expressing $\phi$ in terms of $s=t/T$ we recover
\begin{equation}
\lanran \simeq T^{1/2} \sqrt{\frac{8}{\pi}} \sqrt{s(1-s)}
\end{equation}
exactly as in the continuous case.

In the case of a biased walk, where
$Q(\xi) = (q \delta_{\xi,1} + (1-q)\delta_{\xi,-1})$,
one must replace the number of trajectories $F(x,t)$
with their probability: each trajectory reaching $x$ at time $t$
is weighted by a factor $q^{(t+x)/2} (1-q)^{(t-x)/2}$.
The trajectory leading back to zero has a instead a weight
$q^{(T-t-x)/2} (1-q)^{(T-t+x)/2}$.
The product of these weights gives simply a constant factor $[q(1-q)]^{T/2}$
both in the numerator and in the denominator of formula~(\ref{discretexbm}).
Hence the introduction of a bias does not change all moments of the
distribution $\Omega$ of the excursion.

\subsection{Unbiased Levy flights}
Levy flights are statistical processes of the type~(\ref{bm}),
where the distribution of single steps $Q(\xi)$ has a fat tail
decaying as $|\xi|^{-\mu-1}$ with $0 < \mu < 2$ so that their
variance is infinite.
The standard form of the central limit theorem does not hold
for Levy flights: the invariant distributions under summation
are the Levy stable distributions~\cite{Levy}.
It is therefore natural to wonder whether Levy flights belong to
a different universality class also with respect to the
average excursion.

The analytical evaluation of $\lanran$ for this case is not
straightforward as for the processes discussed so far.
While Eq.~(\ref{xbm}) still holds, the image method cannot
be used to determine $c(x, t \mid x_0,0)$, because the Fokker-Planck equation
for the free process is not local~\cite{Chechkin03}.
We have therefore computed the average excursion numerically,
considering steps performed at discrete integer times
with absolute value distributed according to
$Q(|\xi|) \propto (|\xi|+1)^{-\mu-1}$ and random sign.

Details about the evaluation of $\lanran$ in this and in the
other cases where numerical results have been obtained, are presented
in the Appendix A.

\begin{figure}
\includegraphics[angle=0,width=9cm,clip]{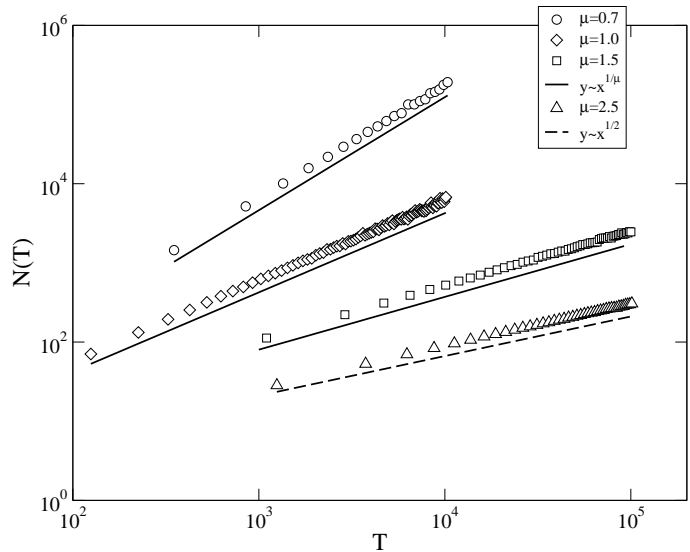}
\caption{
The factor $N(T)$ vs $T$ for Levy flights with several values of $\mu$.
Notice that the exponent is 1/2 for $\mu \geq 2$.}
\label{Fig1}
\end{figure}

To check the validity of the scaling hypothesis we compute the quantity
$N(T)$, such that $\lanran= N(T) f(t/T)$, where we choose, with no
loss of generality to normalize the scaling function $f$ so that
$\int_0^1 f(s) ds=1$. If scaling holds, $N(T)$ has to be proportional to
$T^\alpha$ and the normalized average shapes $\lanran/N(T)$ must collapse
on the same curve for different $T$.
In Fig.~\ref{Fig1} we plot $N(T)$ for several values of $\mu$,
showing that $\alpha=\max[1/2,1/\mu]$.
Also in this case $\alpha$ coincides with the wandering exponent of the
free process, which is $1/\mu$ for $\mu < 2$ and is the usual
diffusive one when $\mu \geq 2$~\cite{Ding95}.

\begin{figure}
\includegraphics[angle=0,width=9cm,clip]{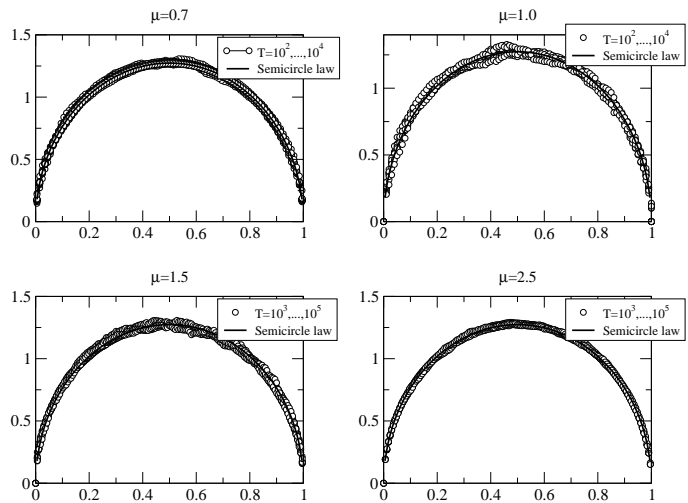}
\caption{
Scaling function $f$ for for Levy flights with several values of $\mu$.}
\label{Fig2} 
\end{figure}

Fig.~\ref{Fig2} reports the shape of the average excursion for
values of $\mu$ such that the second or even the
first moment of the single step distribution is infinite.
In all cases the curves for different values of $T$ collapse
and the scaling form is exactly the same of the Brownian motion:
The shape of the average excursion is completely independent
from the distribution of single steps.

\begin{figure}
\includegraphics[angle=0,width=9cm,clip]{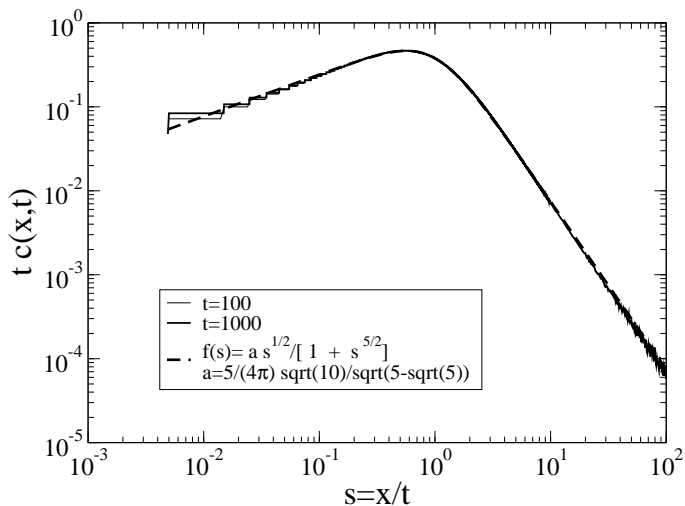}
\caption{
Comparison between $c(x,t)$ evaluated numerically for for Levy flights with steps distributed
according to Eq.~(\ref{Cauchy}) and the ansatz~(\ref{cCauchy}).
}
\label{Fig3} 
\end{figure}

In the Appendix B we report the results also for the variance of Levy flights.
We have not been able to proof this result analytically for Levy flights.
However, in the case $\mu=1$, we have considered the Levy-stable distribution,
the Cauchy distribution
\be
Q(\xi) = {1 \over \pi (1+\xi^2)}.
\label{Cauchy}
\ee
We have computed numerically the probability density $c(x,t)$
for such a process. The result is presented in Fig.~\ref{Fig3},
where it is compared with the ansatz
\be
c(x,t) = {a \over t} {\sqrt{x/t} \over
1 + \left({x/t}\right)^{5/2}},
\label{cCauchy}
\ee
where $a$ is a normalization constant.
Formula~(\ref{cCauchy}) is the simplest expression that interpolates 
between the small $s$ and large $s$ power law behaviors found numerically. 
The agreement between the numerical results and the formula is striking.

Inserting the expression~(\ref{cCauchy}) into Eq.~(\ref{xbm}) and performing
the integrals, one obtains
\be
\lanran = T \sqrt{s(1-s)}.
\ee
adding another piece of evidence to the universality of $f(s)$.

\subsection{Biased Levy flights}

We now consider the case of  biased Levy flights, where
single increments are distributed symmetrically as in the unbiased case
plus a constant term $v$.
In this case the average form of the excursion is in general
asymmetric for finite times 
(toward left or right depending on the sign of $v$).
To understand this behavior, it is fundamental to
consider the relative importance in the equation of motion of the drift
term and of the wandering due to the stochastic variable $\xi$.

The former is clearly $vT$, while the latter grows as $T^{1/\mu}$.
Hence a crossover time $T^*(v) \sim v^{-\mu/(1-\mu)}$ exists between
two regimes: which of the two mechanisms dominates
depends on the value of $\mu$.
For $1<\mu<2$ the bias dominates for large times, while the wandering
is larger than $vT$ for $T \ll T^*$. For $\mu<1$ the opposite is
true and asymptotically the bias does not play any role.

When the bias is irrelevant, the behavior is the same of the unbiased
case: $\lanran$ is given by Eq.~(\ref{semicircle}), $\alpha=1/\mu$ and
the first passage time distribution $P(T)$ decays as
$T^{-1-1/\mu}$~\cite{note2}.

\begin{figure}
\includegraphics[angle=0,width=9cm,clip]{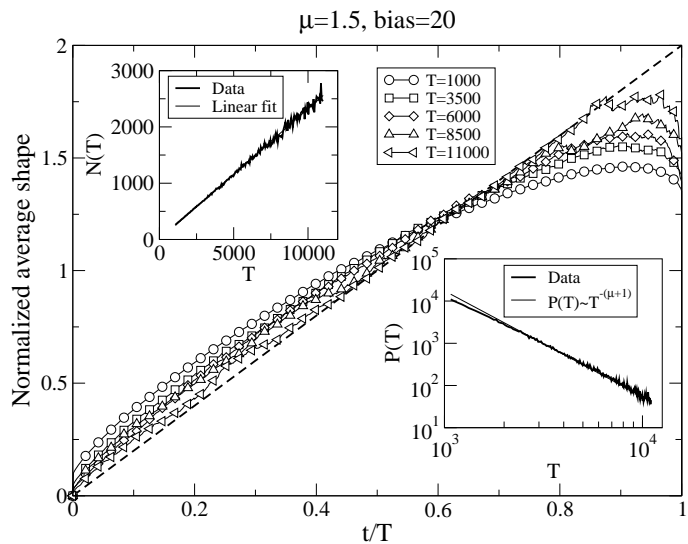}
\caption{
Bias-dominated regime for Levy-distributed increments ($\mu=1.5$)
with bias $v=20$.
Main: Normalized scaling function $f$ converging toward the asymptotic
form $f(s)=2 s$.
Upper inset: The factor $N(T)$ growing as $T^{\alpha}$ with $\alpha=1$.
Lower inset: First return distribution $P(T)$ decaying as $T^{-\mu-1}$.}
\label{Fig4}
\end{figure}

\begin{figure}
\includegraphics[angle=0,width=9cm,clip]{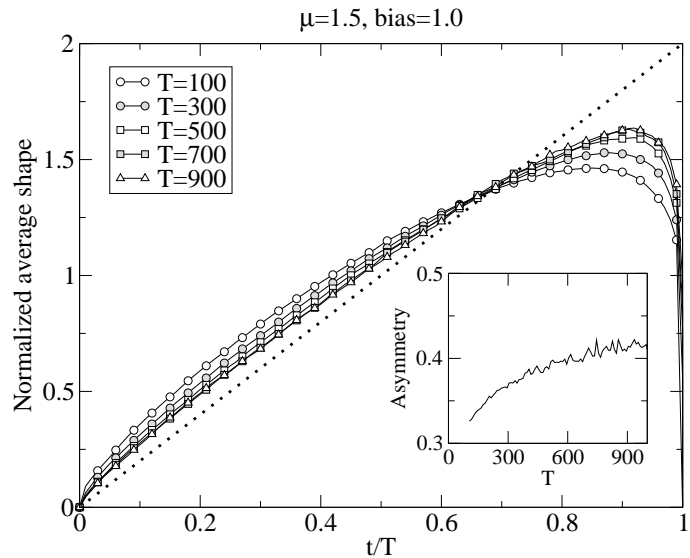}
\caption{Main: Scaling function $f$ for biased Levy flights with $\mu=1.5$ and $v=1$.
Inset: Temporal evolution of the asymmetry parameter
[Eq.~(\ref{asymmetry})].}
\label{Fig5}
\end{figure}

\begin{figure}
\includegraphics[angle=0,width=9cm,clip]{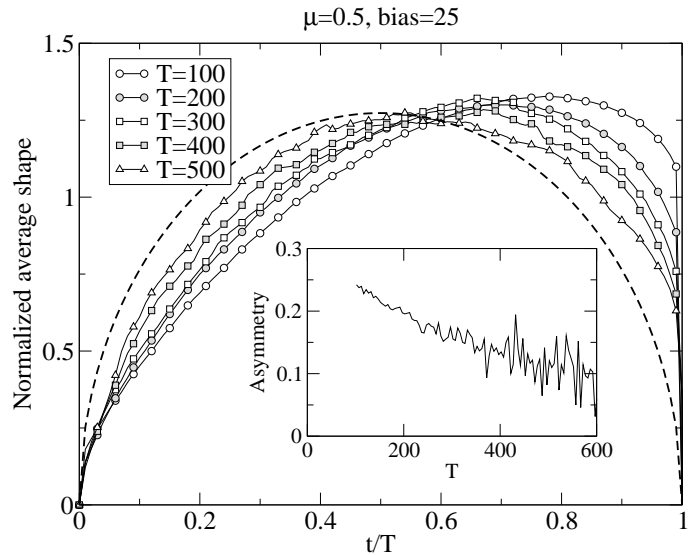}
\caption{Main: Scaling function $f$ for biased Levy flights with $\mu=0.5$ and $v=25$.
Inset: Temporal evolution of the asymmetry parameter 
[Eq.~(\ref{asymmetry})].}
\label{Fig6}
\end{figure}

When the bias dominates, instead, the shape of the excursion
is completely different.  In this case the trajectories are
practically deterministic, i. e. ballistic motions with velocity $v$.
However the noise term is crucial to have the constraint $x(T)=0$
satisfied, since at $t \simeq T$ a very large fluctuation is needed.
As a consequence, the distribution of return times is $P(T) \sim
T^{-1-\mu}$, the exponent $\alpha$ is 1 and the average excursion has
a triangular shape $f_T(s)=2 s$.  This bias-dominated regime is shown
for $\mu=1.5$ and $T \gg T^*$ in Fig.~\ref{Fig4}.
The crossover between the two asymptotic regimes is very slow and it is
not possible to run a single simulation long enough to exhibit the
full transition between the early and late regimes.
Nevertheless it is possible to distinguish clearly between the
behavior for $\mu>1$ and $\mu<1$.
In the former case (Fig.~\ref{Fig5}) the form of the average excursion
becomes more and more skewed with time, while in the latter case
(Fig.~\ref{Fig6}) the opposite behavior is observed.
To give a measure of the asymmetry, we consider the quantity
\be
\int_0^1 ds f(s) \, \, \mbox{sign(s-1/2)} \,
\label{asymmetry}
\ee
which is zero for the semicircle and $1/2$ for the triangle.
The insets of Figs.~\ref{Fig5} and \ref{Fig6} clearly indicate that, for
$\mu=1.5$, it grows, while it decreases to zero for $\mu=0.5$,
consistently with the argument presented above.
 
\begin{figure}
\includegraphics[angle=0,width=9cm,clip]{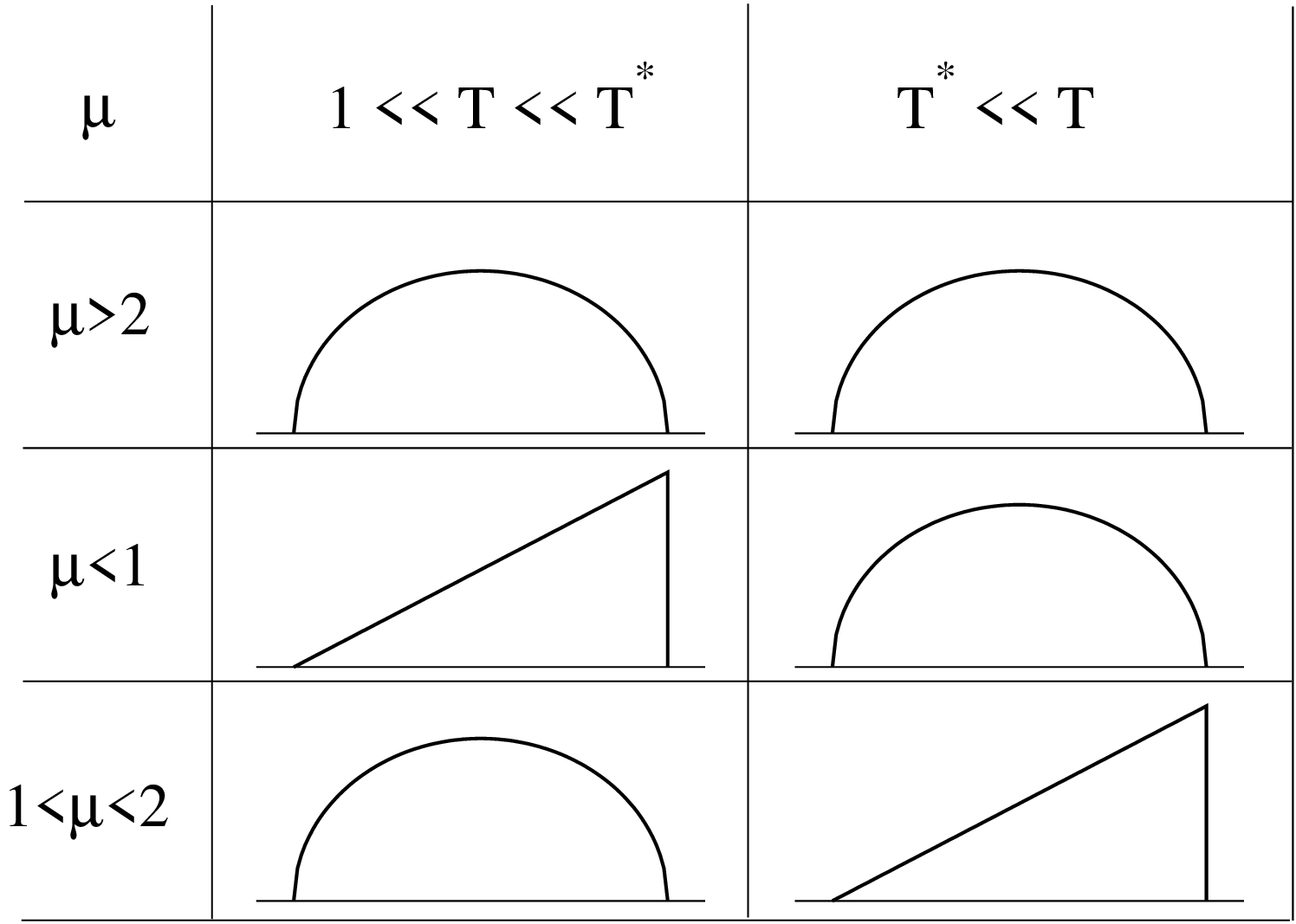}
\caption{Sketch of the average shape for positively biased flights. The
crossover time $T^*$, depending on the value of the bias, changes the
asymptotic behavior only for $1<\mu<2$.  For $\mu<1$ the average
trajectory is asymmetric in the intermediate regime $T \ll T^*$, and it
takes a triangular shape in the limit $T\to \infty$ after
$T^*\to \infty$.}
\label{Fig7}
\end{figure}

Therefore, we can conclude that, while in the unbiased case
the average excursion of a generic uncorrelated stochastic process
with symmetric steps obeys asymptotically the scaling
form~(\ref{scaling}) with universal shape $f_U(s)$ [Eq.~(\ref{semicircle})],
in the biased case, this is true only for $\mu<1$ or $\mu \ge 2$.
For $1< \mu< 2$ instead, the presence of a bias leads to an asymptotic
average excursion with triangular shape.
The results for biased flights are summarized in Fig.~\ref{Fig7}~\cite{note3}.

\section{Damped Brownian motion}
\label{Damped}

We now deal with a Brownian motion in a potential,
Eq.~(\ref{potential}).
We treat only the simplest possible case,
an harmonic potential $V(x) = \lambda x^2$, so that the
Brownian motion is pushed toward the origin by a linear damping term
\be
\partial_t x(t)= - \lambda x + \xi(t)
\label{dbm}
\ee

The formal solution of this equation is given by 
\be
x(t)=x_0 e^{-\lambda t} +\int_0^t ds e^{-\lambda(t-s)} \eta(s)
\ee
Since $x(t)$ is linearly related to $\eta(t)$, it has a Gaussian 
distribution characterized by its first and second moment, that can be 
obtained by averaging over the noise: 
\be
\begin{array}l
m(t)=\langle x(t)\rangle=x_0 e^{-\lambda t} \\
\sigma_{\lambda}^2(t)=\langle (x(t)-\langle x(t)\rangle)^2\rangle=
\frac{1-e^{-2\lambda t}}{2 \lambda}.
\end{array}
\ee
The normalized probability density for the free process is therefore 
\be
P(x,t)=\frac{1}{\sqrt{2\pi}\sigma_{\lambda}(t)} 
\exp{\left[\frac{(x-x_0 e^{-\lambda t})^2}{2 \sigma_{\lambda}^2(t)}\right]}.
\ee
One can easily check that $P(x,t)$ solves the Fokker-Planck equation 
\begin{equation}
\partial_t P(x,t)=\frac{1}{2} \partial^2_x P(x,t)
-\lambda \partial_x [x P(x,t)]
\label{fpdbm}
\end{equation}
associated to the Langevin equation (\ref{dbm}).

The distribution of first return times is
\be
P(T) \propto 
\frac{1}{8 \pi} \frac{e^{2\lambda T}}{\sigma_{-\lambda}^{3}(T)}
=\frac{1}{8 \pi}(2\lambda)^{3/2}
\frac{e^{-\lambda T}}{(1-e^{-2\lambda T})^{3/2}}
\,.
\ee
Thus 
\be
P(T)\propto \left\{
\begin{array}{ll}
{\frac{8}{\pi}}T^{-3/2}  & T \ll 1/\lambda \\
{\frac{8}{\pi}}(2\lambda)^{3/2}e^{-\lambda T} & T \gg 1/\lambda \,,
\end{array}
\right.
\ee

For $T \ll 1/\lambda$, the $T^{-3/2}$ behavior characteristic of 
Brownian motion is recovered. When $T$ is of order $1/\lambda$,
the power law behavior is cut off exponentially.

To calculate the average excursion we need to evaluate 
the probability $c(x,t \mid x_0,0)$ that a walk originating at $x_0$
at time $0$ is found in $x$ at time $t$, without having ever touched 
the origin.
Since Equation~(\ref{dbm}) is linear, the image method can
be applied, yielding
\be
c(x,t \mid x_0,0) =
\frac{1}{\sqrt{2 \pi} \sigma_{\lambda}(t)}
\left\{
\exp{\left[
-\frac
{\left(x- x_0 e^{-\lambda t}\right)^2}
{2 \sigma^2{_\lambda}(t)}
\right]} -
\exp{\left[
-\frac
{\left(x+ x_0 e^{-\lambda t}\right)^2}
{2 \sigma^2{_\lambda}(t)}
\right]} 
\right\} \,.
\ee
In the limit $x_0 \rightarrow 0$ 
\begin{equation}
c(x,t \mid x_0,0) \propto
\frac{2 x_0 x e^{-\lambda t}}{\sqrt{2 \pi} \sigma_\lambda^{3}(t)}
\,\,
e^{-\frac{x^2}{2\,\sigma^2_\lambda(t)}}.
\label{f+}
\end{equation}

The quantity $c(x_0,T \mid x ,t)$ for $x_0 \rightarrow 0$ is obtained
in a similar way
\begin{equation}
c(x_0,T \mid x,t) \propto
\frac{2 x_0 x e^{-\lambda (T-t)}}{\sqrt{2 \pi} \sigma_\lambda^{3}(T-t)}
\,\,
e^{-\frac{x^2  e^{-2\lambda(T-t)}}{2\,\sigma^2_\lambda(T-t)}}.
\label{f-}
\end{equation}
Defining $\tilde{\sigma}^2_\lambda(t)=\sigma^2_\lambda(t) e^{2\lambda t}$ 
one can rewrite (\ref{f-}) as
\begin{equation}
c(x_0,T \mid x,t) \propto 
\frac{2 x_0 x e^{2\lambda (T-t)}}{\sqrt{2 \pi}
\tilde{\sigma}_\lambda^{3}(T-t)}
\,\,
e^{-\frac{x^2}{2\,\tilde{\sigma}^2_\lambda(T-t)}}.
\end{equation}
Note that
$\tilde{\sigma}^2_\lambda(t)=\sigma^2_\lambda(t) e^{2\lambda t}
=\sigma^2_{-\lambda}(t)$, thus the process with reversed time
formally corresponds to the process with $\lambda \rightarrow -\lambda$.

The distribution of excursions $\Omega$ is then
\be
\Omega(x,t \mid x_0,0;x_0,T) \propto
e^{-\lambda T} [\sigma_\lambda(t) \sigma_{-\lambda}(T-t)]^{-3}
e^{-x^2/[2 \sigma^2_{eq}(t,T)]} \,,
\ee
where the variance of the Gaussian factor is
\begin{equation}
\sigma^2_{eq}=(\sigma_{\lambda}^{-2}(t)+ 
\sigma_{-\lambda}^{-2}(T-t))^{-1}=\frac{1}{2\lambda}
\left[
\frac{(1-e^{-2\lambda t})(1-e^{-2\lambda(T-t)})}{1-e^{-2\lambda T}}
\right]. 
\end{equation}

Inserting Eqs.~(\ref{f+}) and~(\ref{f-}) into Eq.~(\ref{xbm}) yields
\begin{equation}
\lanran
= \sqrt{\frac{8}{\pi}} \sigma_{eq}(t,T)
= \sqrt{\frac{8}{\pi}}
\frac{1}{\sqrt{2 \lambda}}
\sqrt{
\frac{(1-e^{-2\lambda t})(1-e^{-2\lambda(T-t)})}
{1-e^{-2\lambda T}}}.
\label{damped}
\end{equation}
\begin{figure}

\includegraphics[angle=-90,width=9cm,clip]{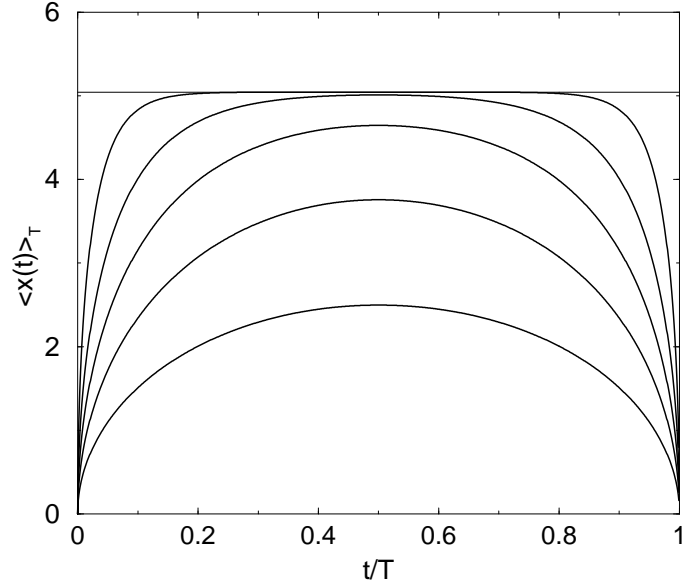}
\caption{Average excursion for a damped random walk
[Eq.~(\ref{damped})] for $1/\lambda=20$. From top to bottom lines are
for $T=10, 25, 50, 100, 250$. Notice that the shape flattens to the
constant value $\sqrt{4/(\pi \lambda)}$ (thin line).}
\label{Fig8}
\end{figure}

As expected the existence of a characteristic time in the problem
($1/\lambda$) breaks down scaling; the shape of
the average excursion changes with $T$ (Fig.~\ref{Fig8}).
However formula~(\ref{scaling}) still holds in the two asymptotic
limits
\begin{equation}
\lanran
=\left\{
\begin{array}{ll}
\sqrt{\frac{8}{\pi}}\sqrt{\frac{t(T-t)}{T}}  & t,T-t \ll 1/\lambda \\
\sqrt{\frac{8}{\pi}}{\frac{1}{\sqrt{2\lambda}}} & t,T-t \gg 1/\lambda \,.
\end{array}
\right.
\end{equation}
Thus in the small $\lambda$ limit the semicircle is recovered, while 
in the large $\lambda$ limit the curve flattens around a value 
proportional to $1/\sqrt{\lambda}$, while keeping the semicircular tails.
The crossover between the two regimes corresponds to a change in the
variance $\sigma^2_{eq}$ of $\Omega$
\be
\sigma^2_{eq}=\left\{
\begin{array}{ll}
\frac{t(T-t)}{T}  & t,T-t \ll 1/\lambda \\
\frac{1}{{2\lambda}} & t,T-t \gg 1/\lambda \, ,
\end{array}
\right.
\ee
Thus fluctuations saturate at the value $1/(2 \lambda)$, indicating 
the absence of correlations on time scales longer than the characteristic 
time $1/\lambda$.

In the computation presented here we have taken the reference value
to coincide with the origin, i. e. $a=0$.
In the presence of a potential, the effect of a reference value
different from zero corresponds to the change $V(x) \rightarrow V(x+a)$.
For the damped Brownian motion, this is equivalent to the introduction
of a bias $\lambda a$.
In general, such a bias perturbs the form of the average excursion, at
odds with the case of free random walks, which are unchanged by the
presence of a bias.
However the change in the average shape is small provided $a$ is small
compared with $4/\sqrt{\pi \lambda}$, the maximal amplitude of $\lanran$.

\section{Long-ranged correlations}
\label{Long-range}

We now start considering the effect of the introduction of temporal
correlations in the process.
In particular, we study processes of the form
$\partial_t x(t) = \xi(t)$, with correlations between single increments
$g(t,t') \equiv \langle \xi(t) \xi(t') \rangle - \langle \xi(t) \rangle
\langle \xi(t') \rangle \neq \delta_{t,t'}$.

We first focus on a process $\partial_t x(t) = \xi(t)$ where the
noise performs in its turn a Brownian motion $\partial_t \xi(t) = \eta(t)$
with $\langle \eta \rangle = 0$ and $\langle \eta^2 \rangle = 1$.
Clearly this process is non-Markovian and can be written as
the Random Accelerated Particle (RAP)
\be
\partial^2_t x(t) = \eta(t).
\ee
The correlation function of $\xi(t)$, $g(t,t')= \min(t,t')$ does not
decay to zero when $t-t'$ diverges: the noise has then infinitely-ranged
correlations.
This process has been studied recently with regards to
polymers~\cite{Gompper89} and the inelastic
collapse of granular matter~\cite{Cornell98}.

It is important to stress here that, at odds with the previous cases,
the non-Markovian nature of the process implies that to define completely
a fluctuation one has also to consider the initial and final velocities
(the process is Markovian if one considers the broader space of 
coordinate and velocity).
We consider separately avalanches beginning with zero or a finite
velocity, that will yield different results.
For what concerns the final velocity the most natural thing is to average
over all final velocities.
The condition on the initial velocity is numerically very easy to implement,
since it corresponds to setting $\eta(0)=v_0$.
Instead, the condition on the final velocity is more delicate.
Considering returns within a strip ($-\epsilon,+\epsilon$)
implies, when using discrete times, a hidden condition on the final velocities.
This is apparent if one considers the long time decay of the distribution
of first return times.
If one considers discrete times and a return in a strip, the known result
$P(T) \sim T^{-5/4}$~\cite{Sinai92} is not recovered.
The expected exponent is found instead if one averages over all
trajectories positive up to $t<T$ and becoming negative for $t=T$.
Therefore we considered only this latter kind of trajectories.
We believe that, asymptotically, this coincides with the continuous time
process averaging over all final velocities.

\begin{figure}
\includegraphics[angle=0,width=9cm,clip]{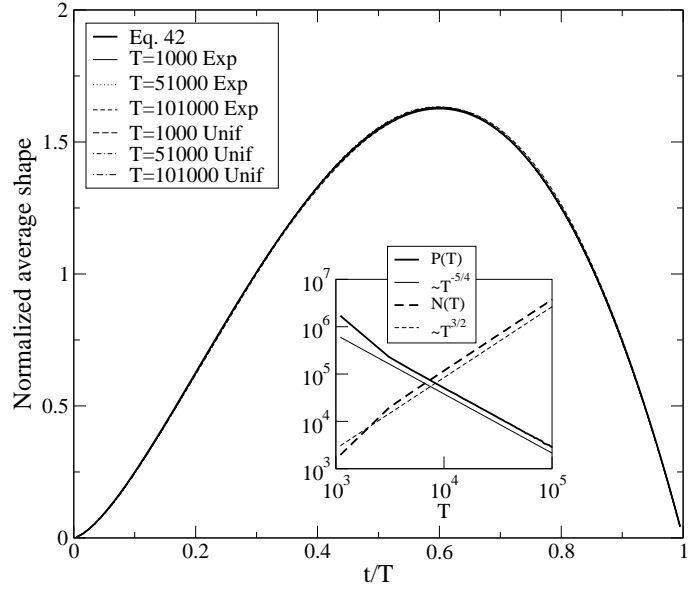}
\caption{
Main: Average excursion for a RAP with zero initial velocity, with uniform
or exponential
distribution of the noise $\eta$, showing perfect agreement with the simple
form~(\ref{Beta}).
Inset: Factor $N(T)$ and first-return time distribution $P(T)$ for uniform
noise distribution.
}
\label{Fig9}
\end{figure}

Figure~\ref{Fig9} reports the average excursion shape for two different
distributions of $\eta$ with finite variance, one uniform and the other
exponential, both in the case of zero initial velocity.
It is clear that scaling holds very well for all values of $T$
considered and for both distributions of the noise, and that the scaling
function is asymmetric.
As shown in the inset, the exponent $\alpha$ is again equal to the
free wandering exponent, which is $3/2$.
Remarkably, the scaling function is indistinguishable from the simple form
\be
f(s)={\frac{3}{2}s^{3/2}(1-s) }\,.
\label{Beta}
\ee

In the Appendix B we report also the results for the variance of the excursion,
showing that it scales as the first moment.

\begin{figure}
\includegraphics[angle=0,width=9cm,clip]{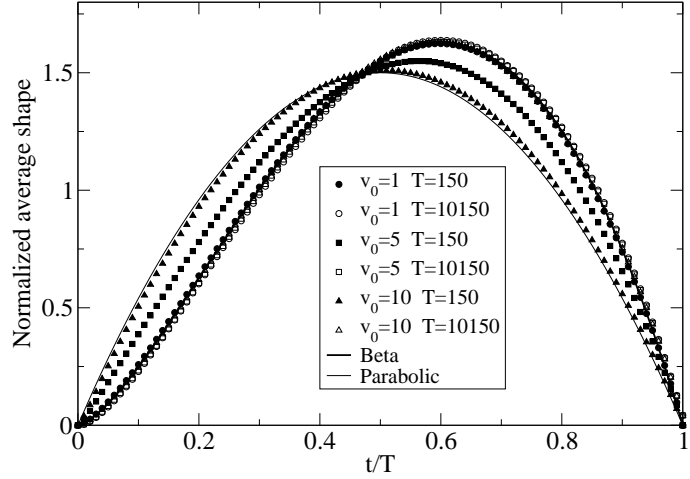}
\caption{
Main: Scaling function $f$ for a RAP with finite initial velocity,
with uniform distribution of the noise $\eta$.}
\label{Fig10}
\end{figure}

When the initial velocity $v_0$ is finite, instead, the asymptotic scaling
associated with an average shape~(\ref{Beta}) is preceded by a transient
regime with a more symmetric $\lanran$, depending on $T$ and $v_0$
(Fig.~\ref{Fig10}).
For large $v_0$ the shape is very close to a simple parabola $6s(1-s)$.
We have also checked that the exponent $\alpha$ crosses over from a
ballistic value $\alpha=1$ at short times to the asymptotic value $3/2$.
Comparing the values of $\alpha$ in the two regimes we expect the crossover
time to be proportional to $v_0^2$.

\begin{figure}
\includegraphics[angle=0,width=9cm,clip]{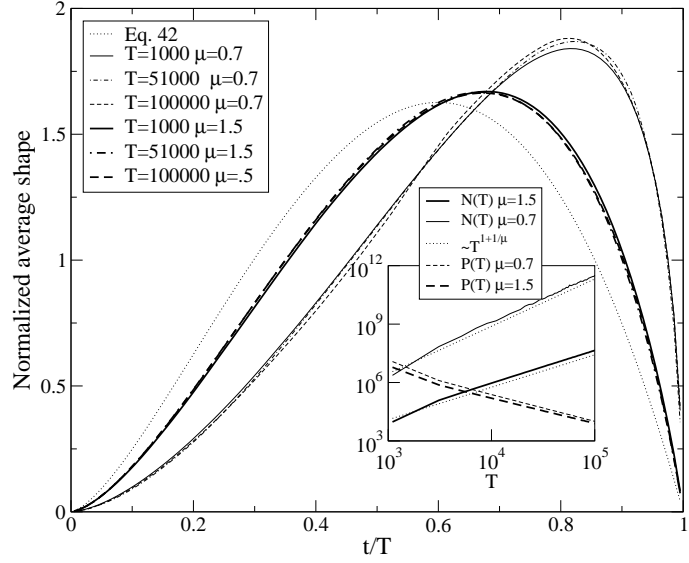}
\caption{
Main: Scaling function $f$ for for a RAP with noise $\eta$ distributed with 
slow decaying tails $P(\mid \eta \mid)$ and several values of $\mu$,
compared with the simple form~(\ref{Beta}).
Inset: Factor $N(T)$ and first-return time distribution $P(T)$ for
the same values of $\mu$.}
\label{Fig11}
\end{figure}

We then consider RAP with noise $\eta$ distributed with slowly decaying tails
$P(|\eta|) \sim \eta^{-1-\mu}$.
With zero initial velocity the scaling function depends on the
value of $\mu$~(Figure~\ref{Fig11}),
at odds with what occurs for uncorrelated unbiased processes.
The skewness is toward right, the more so for small values $\mu$.
The exponent $\alpha$ is equal to $1+1/\mu$.

\begin{figure}
\includegraphics[angle=0,width=9cm,clip]{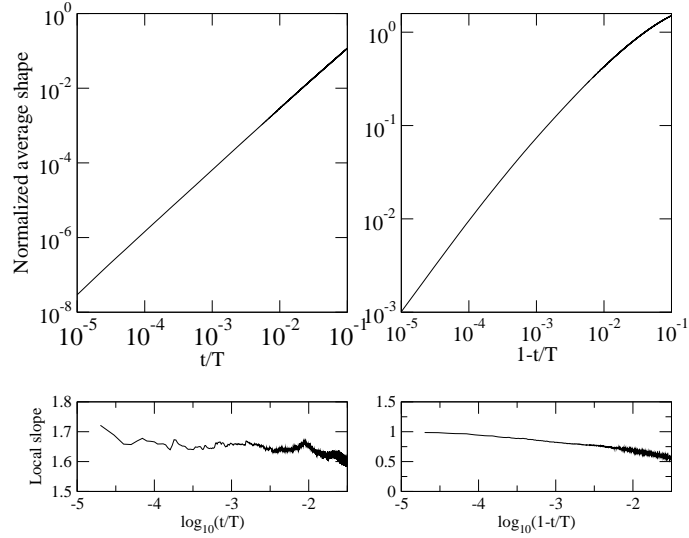}
\caption{
Left top: Small $t/T$ tail of the scaling function $f$
for RAP with Cauchy distribution of single steps. Left bottom:
Local effective exponent computed on the figure above.
Right top: Small $1-t/T$ tail of the scaling function $f$
for RAP with Cauchy distribution of single steps. Right bottom:
Local effective exponent computed on the figure above.
}
\label{Fig12}
\end{figure}

Despite the dependence of the detailed scaling form of $\lanran$ on
the distribution of single steps, some degree of universality remains
also for the RAP process.
The right panel of Figure~\ref{Fig12} shows that the exponent
characterizing the behavior of $f(s)$ for $s \to 1$ is universal,
being linear also for $\mu=1$.
For small values of $s$ instead, a careful analysis indicates that the 
actual exponent of the power-law behavior is slightly different from the value
$3/2$ obtained for finite variance (Figure~\ref{Fig12}, left panel).
We do not have an explanation for the value of this exponent.
However, we cannot rule out that such universality is restored for
larger values of $T$.
We have checked that, in the case of finite initial velocity $v_0$,
the transient parabolic shapes are present also for values of $\mu < 2$.

\section{Short-ranged correlations}
\label{Short-range}

We now turn to the case of short-ranged correlations.
We consider a process with correlations decaying exponentially
over an interval $\tau$
\be
g(t,t')=\exp{(-|t-t'|/\tau)}.
\ee
A process of this type can be obtained in practice by feeding
the random walk with noise obeying a damped random walk
of the form
$\xi(t+1)=\gamma \xi(t) + (\sqrt{1-\gamma^2})\eta(t)$
with uncorrelated $\eta$.
The correlator of $\xi(t)$ is easily shown to decay exponentially,
with a characteristic time $\tau = -1/\log \gamma$.

For such a process, we do not expect scaling~(\ref{scaling})
to hold for all values of $T$. We can anticipate instead two regimes
depending on the duration $T$ of the trajectories considered.
For short times $T \ll \tau$, noise is correlated during the
whole trajectory and the behavior must be the same of the case
with infinitely-ranged correlations treated above.
For long times $T \gg \tau$, on the contrary, noise is
correlated only for intervals that are short compared to the
total duration of the excursion. Hence the process is equivalent
to an uncorrelated process with some effective distribution of the
single increments.

\begin{figure}
\includegraphics[angle=0,width=9cm,clip]{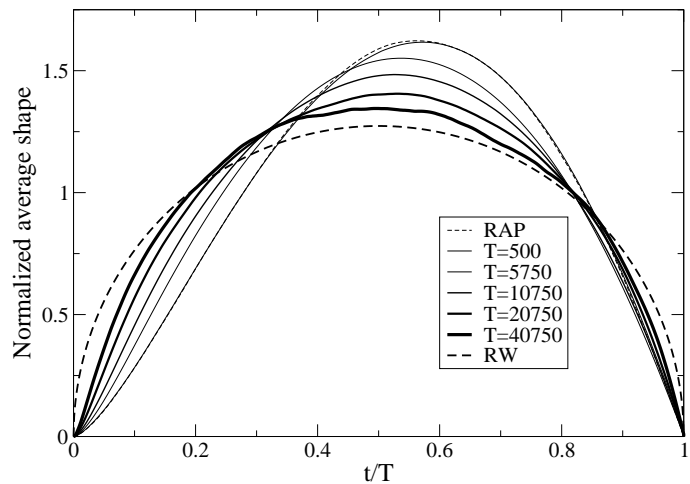}
\caption{
Normalized average shape for a short-memory process
with zero initial velocity, finite variance noise
and $\tau=1000$. The dashed lines are the expected limiting curves
for $T \ll \tau$ and $T \gg \tau$.
}
\label{Fig13}
\end{figure}

Numerical results fully confirm this picture.
In Fig.~\ref{Fig13} we show the case of a short-memory process
with zero initial velocity and finite variance noise.
For short times $T \ll \tau$ the shape is very close to the
form~(\ref{Beta}), valid for RAP.
A slow crossover leads for longer times to the semicircle law
valid for uncorrelated processes.
We have checked that the expected pattern of behavior occurs also
for non zero initial velocity and for Levy-distributed noise.

\section{Conclusions}
\label{Conclusions}

Let us summarize the results presented in the previous sections.  We
have studied the statistics of excursions in some classes of
stochastic processes, with particular attention to the average shape.
For uncorrelated free processes we have found that the average
excursion has a scaling function proportional to a semicircle,
independent from the distribution of the single steps provided it is 
symmetric.  This holds not
only for distributions that renormalize to the Gaussian, but also for
the class of distributions that renormalize to symmetric
Levy stable distributions. More generally the scaling
function is unchanged when a bias is introduced, with the notable
exception of the case with $1 <\mu <2$, where the asymptotic shape of
fluctuations is triangular.  The addition of a linear damping term in
the Langevin equation for the process introduces a characteristic
time scale, that separates between two regimes: for short times the
process is dominated by noise, and the excursion is the same as in the
free case.  For longer times, scaling breaks down and the shape of the
average fluctuation flattens to a value independent from its duration.
Furthermore, we have analyzed the effect of noise correlations for the
free process. When correlations are long ranged the shape of a
fluctuation depends on the initial velocity $v_0$. For $v_0=0$ we find
that the scaling function has asymmetric tails $s^{3/2}$ and $(1-s)$
in the Gaussian case, while the situation for Levy distributed steps
is less clear. For $v_0 >0$ a transient regime exists such that the
scaling function has linear tails (independent from the distribution
of the single steps), before it crosses over to the asymptotic form
which is the same of the $v_0=0$ case.  Finally, in the case of short
range correlated noise, the range of correlation sets a time scale
that separates between a short time regime, where, as expected the
behavior is similar to the long range case, followed by a crossover
to the asymptotic uncorrelated behavior.

Application of this analysis to real data requires some care.
Indeed in many situations of practical interest one deals
with long time series consisting of a large number of successive
fluctuations. 
In such a case, if one computes $\lanran$ by averaging
over successive returns to the value $x=a$ one may average over pulses
that are not statistically independent. 

In our work, on the contrary, we take care to average always on
independent events.  When the process we consider is Markovian this
does not require particular prescriptions. In this case averaging over
successive fluctuations in a single realization is equivalent to
averaging over avalanches belonging to different realizations.
Otherwise one should consider avalanches separated by times
larger than the largest correlation time in the system, if the
analysis is restricted to a single realization.  If the correlation 
time is infinite, as in the RAP, one should consider only fluctuations
belonging to independent realizations.  

Another relevant issue for the application to real time series
concerns the amount of events required to obtain sufficiently clean
results.  In principle one should average over fluctuations of exactly
the same duration, and rescale afterward. This may turn out to require
an exceedingly large number of events. An alternative procedure is to
assume scaling and average over fluctuations of different duration,
properly rescaled, with an exponent that can be obtained by plotting
the size of fluctuations as a function of their duration.  This also
checks whether scaling holds or not.

For the case of Barkhausen noise, which was the initial inspiration of
this work, we suspect  that the asymmetric shape observed in experiments
must be due to the presence of some kind of correlations.  However,
the kind of correlations that we have analyzed give rightward asymmetric
shapes, while the one observed experimentally are leftward.
This calls for further analysis, of more general processes.

\appendix
\section{Details about the numerical results}
\label{Appendix1}

When performing numerical simulations, we have taken times to be
discrete and space continuous, thus the concept of first return to the
initial value needs some clarification.  We have considered the
process to return to the initial value when its value is in a small
interval around it $[-\epsilon, \epsilon]$.  In order not to introduce
an artificial asymmetry we have applied the same condition to the
first step of the excursion as well.  This is implemented by letting
the process start at $x=0$ for some negative time and taking as $t=1$
the first time such that $x(t)> \epsilon$.  The average is then
performed over all trajectories that first return between $-\epsilon$
and $\epsilon$ at a specified time $T$ under the constraint that
$x(t)> \epsilon$ for $1 < t < T$.  Care has to be used when choosing the
value of $\epsilon$, which should be as small as possible to
give $\epsilon$ independent results. At the same time too small values
of $\epsilon$ make the numerical simulation very time consuming.

In all simulations $T$ is integer; hence it is in principle
possible to average only over trajectories that return after
{\em exactly} $T$ steps.
However $T$ cannot be decided a priori but is the
outcome of the simulation; collecting a large number of
trajectories for a single large $T$ may be a prohibitive task.
Therefore, similarly to what is done in experiments,
we have performed a binning procedure, by averaging over
all runs that return within a time interval between $T-\Delta T$ and
$T+\Delta T$.
Since the relative size of the bins $2 \Delta T/T$ decreases as $T$ grows, we
are sure that this procedure does not lead to significant perturbation
of the results for large $T$.

Finally, when presenting numerical evaluations of the average excursion we
plot $\lanran$ normalized by the factor
$N(T)= \int_0^1 ds \langle x(sT) \rangle _T$.
This allows to check the form~(\ref{scaling}) in two ways.
If scaling holds then $N(T)$ grows as a power-law (the exponent is
$\alpha$) and the different curves collapse on a universal
shape [which is $f(s)$].

\section{Numerical computation of variance}

\begin{figure}
\includegraphics[angle=0,width=9cm,clip]{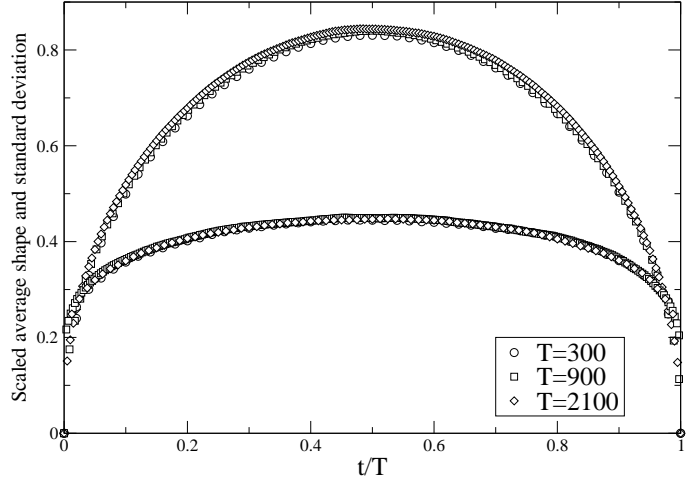}
\caption{
Plot of the average shape and standard deviation for Levy flight with $\mu=1.5$.
Curves for different durations are divided by the scaling factor $T^{1/\mu}$.
The upper curves are the average shapes, the lower ones are the standard deviations.}
\label{Fig14}
\end{figure}

\begin{figure}
\includegraphics[angle=0,width=9cm,clip]{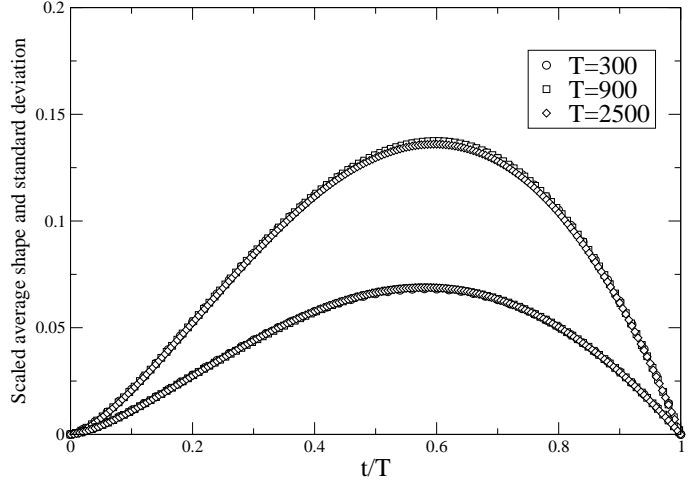}
\caption{
Plot of the average shape and standard deviation for RAP with Gaussian noise.
Curves for different durations are divided by the scaling factor $T^{3/2}$.
The upper curves are the average shapes, the lower ones are the standard deviations.}
\label{Fig15}
\end{figure}

For the simplest processes that can be attacked analytically we have
already computed the variance:
\[
\sigma^2_T(t)=\langle (x(t)-\langle x(t)\rangle)^2\rangle
\]
In the other cases it is possible to determine numerically such a
quantity. In this appendix we show, as an example, the results for two of such cases,
i.e. the unbiased Levy flight with $\mu=1.5$ and the RAP.
In Figures~\ref{Fig14} and~\ref{Fig15} we have reported the average excursion
$\lanran$ for different durations $T$ divided by the expected scaling factor $T^\alpha$.
Correspondingly we plot the values of the standard deviation $\sigma_T(t)$ divided by
the same factor.
One sees that $\sigma$ scales exactly as the average shape in both cases.

\begin{thebibliography}{99}
\bibitem{note1}
In this case the application of the image method is less trivial
than in the unbiased case. The function $c$ is the sum of a
Gaussian propagating with velocity $v$ and an anti-Gaussian,
propagating with the same velocity $v$ but with a weight
$e^{2 x_0 v}$ (see Ref.~\cite{Rednerbook} for details).
\bibitem{note2} Note that $P(T)$ is the probability for the walk to be
positive up time $T$ and in an $\epsilon$ interval around the origin
at time $T$. This is different from the probability of a positive walk
up to time $T$ that becomes non positive at time $T$, which decays as
$T^{1/2}$, independently from $\mu$.  See Ref.~\cite{Feller}.
\end{thebibliography}
\end{document}